%% file: manuscript.tex
\documentclass[a4paper,fleqn,usenatbib]{mnras}

\usepackage{newtxtext,newtxmath}

\usepackage[T1]{fontenc}
\usepackage{ae,aecompl}

\input{packageIncludes}

\newcommand{\msun}{\ensuremath{\mathrm{M_{\odot}}}}

\input{figureMacros}

\DeclareSIUnit[number-unit-product = \;]
\solarmass{\msun}

\DeclareSIUnit[number-unit-product = \;]
\dyn{dyn}

\DeclareSIUnit[number-unit-product = \;]
\ergy{erg}

\hypersetup{
    colorlinks,
    citecolor=blue,
    filecolor=black,
    linkcolor=blue,
    urlcolor=black
}

\newcommand{\Mrms}{\mathcal{M}_\mathrm{rms}}
\newcommand{\urho}{\mathrm{g\,cm^{-3}}}
\newcommand{\uT}{\mathrm{K}}
\newcommand{\uEktd}{\mathrm{erg\,cm^{-3}\,s^{-1}}}
\newcommand{\ut}{\mathrm{ms}}
\newcommand{\uvkms}{\mathrm{km\,s^{-1}}}
\newcommand{\es}[2]{#1\times10^{#2}}
\title[SN Ia DDT Explosion Mechanism]
{SN Ia DDT Explosions Powered by the Zel'dovich Reactivity Gradient Mechanism}
\author[Brooker, Plewa, \& Fenn]
{E. Brooker$^{1}$\thanks{E-mail: eb11d@my.fsu.edu (EB); tplewa@fsu.edu (TP); fenn1@llnl.gov (DF)},
 T. Plewa$^{1\star}$
 and
 D. Fenn$^{2\star}$\\
  $^1$Department of Scientific Computing, Florida State University, Tallahassee, FL 32306, U.S.A.,\\
  $^2$Lawrence Livermore National Laboratory, Livermore, CA, 94550 U.S.A.}
\date{Accepted 2020 August 11. Received 2020 August 8; in original form 2020 May 21}
\pubyear{2020}
\begin{document}
\pagerange{{000}--{000}} \pubyear{2020} \volume{000}
\label{firstpage}
\maketitle
%
%
%
%
%
%
\begin{abstract}
The aim of this work is to identify and explain the necessary conditions required for an energetic explosion of a Chandrasekhar-mass white dwarf.
We construct and analyze weakly compressible turbulence models with nuclear burning effects for carbon/oxygen plasma at a density expected for deflagration-to-detonation transition (DDT) to occur.
We observe formation of carbon deflagrations and transient carbon detonations at early times. As turbulence becomes increasingly inhomogeneous, sustained carbon detonations are initiated by the Zel'dovich reactivity gradient mechanism. The fuel is suitably preconditioned by the action of compressive turbulent modes with wavelength comparable to the size of resolved turbulent eddies; no acoustic wave is involved in this process. Oxygen detonations are initiated either aided by reactivity gradients or by collisions of carbon detonations.
The observed evolutionary timescales are found sufficiently short for the above process to occur in the expanding, centrally ignited massive white dwarf. The inhomogeneous conditions produced prior to DDT might be of consequence for the chemical composition of the outer ejecta regions of SN Ia from the single degenerate channel, and offer potential for validation of the proposed model.
\end{abstract}
%
%
%
%
%
%
%
%
\section{Introduction}\label{s:intro}
The deflagration-to-detonation transition (DDT) mechanism remains one of the major unsolved problems of theoretical and computational combustion. It has been directly observed in a number of laboratory experiments and has been extensively studied by means of computer simulations \citep[][and references therein]{Oran2007}. In the context of astrophysics and stellar evolution, it has now been suspected for almost 30 years that DDT is directly responsible for at least a subclass of white dwarf explosions responsible for luminous Type Ia supernovae (SN Ia). In this case, however, evidence for the DDT is only indirect, relying on post-explosion observational data and speculative physics models such as delayed-detonation \citep[DD;][]{Hoeflich1995}. This particular early study was suggestive of DDT occurring at relatively low densities, $\varrho_\mathrm{tr}\approx 2\times 10^7$ g/cc, which was quickly linked to a morphological change from the flamelet to stirred flame regime \citep[FSF;][]{Niemeyer+97,Khokhlov+97} expected of centrally-ignited deflagrations. Then, as theorized by \cite{Khokhlov1991b}, a DDT might be possible provided that (Rayleigh-Taylor-driven and flame-generated) turbulent perturbations satisfy certain minimal amplitude and size requirements.

The above basic DD explosion scenario was refined in the course of major research efforts. S. Woosley and collaborators used a one-dimensional model of a turbulent flame to study the flame evolution in various burning regimes, including the FSF transition. They provided firmer constraints on conditions for DDT, speculated about possible collective effects in multidimensions \citep{Woosley+09}, and assessed the role of realistic composition \citep{Woosley+11}. \cite{Schmidt+10} adopted a probabilistic DDT framework originally developed by \cite{Pan+08} to account for intermittency of turbulence as a possible source of large-scale fluctuations at the flame front. They provided an independent set of DDT constraints and discussed their dependence on the adopted intermittency model. More recently \citet{Poludnenko2019} postulated a similarity between the terrestrial and astrophysical DDT mechanisms. In that scheme, interaction between a deflagration with nearly incompressible turbulence produces a shock and ultimately results in a detonation. However, neither this nor other briefly discussed here efforts demonstrated DDT to operate under conditions expected for SNe Ia.

In this work, and in contrast to the previous SN Ia DDT studies, we consider a situation in which no flame is initially present, nuclear burning is solely due to self-heating, and turbulence is created and sustained in a fully controlled manner. The use of a minimal set of physics makes our results free of numerous assumptions present in other models. This also makes the presented results easily reproducible as the required computational model is simple and can be built using readily available software packages
\section{Methods and initial conditions}\label{s:methods}
We performed computer simulations using {\sc Proteus}, a fork of the {\sc FLASH} code \citep{Fryxell+00}. We used the PPM method \citep{ColWood84} to solve the hydrodynamic equations of stellar plasma with thermodynamics described by the Helmholtz equation of state \citep{TimmesSwesty2000}. The hydrodynamics solver was supplemented by equations for advection of nuclear species and nuclear burning source terms for energy and species evolution due to nuclear reactions. In the course of trial simulations, we found that the results obtained with a small nuclear network of 7 species, {\sc Iso7}, did not differ in any appreciable way from the results obtained with larger (but more expensive) 19 isotope network, {\sc Aprox19}. Consequently, we adopted {\sc Iso7} for the remaining set of simulations.

Simulations were performed on periodic domains with $32\ \mathrm{km}$ length on each side. Because of our interest in DDT, we assumed an initial plasma density of $\rho_{0} = \es{1}{7}\ \urho$ and plasma temperature of $T_{0} = \es{1}{9}\ \uT$, with a 50/50 carbon/oxygen composition. This particular choice of the initial conditions, especially density slightly lower than the observationally-calibrated DDT value of $\rho_{tr} = \es{2}{7}\ \urho$ \citep{Hoeflich1995}, does not affect major conclusions of this work. 

We produced a turbulent quasi-steady state using the spectral forcing method \citep{EswaranPope1988,Federrath2010} to allow for compressibility in the drive. The forcing spectrum contained $50$ per cent of energy in compressible modes and was flat in shape, extending from $k=1$ to $k=4$. The individual modes were assumed to decay on a time-scale of $10\ \ut$ while the spectrum shape was updated in time with resolution of $0.1\ \ut$. The evolution towards quasi-steady state was performed over the initial $75\,\ut$ of simulations, which amounts to about three turbulent eddy turnover times. During that transient stage, the total internal energy in the simulations was kept constant \citep{FennPlewa2017}. 

Because we were interested in the dependence of the model outcomes on the characteristic turbulence Mach number in the weakly compressible regime, we systematically varied the turbulence forcing kinetic energy in the range $E_{k,td} = \es{(1-1.5)}{15}\ \uEktd$. For the model with $\Mrms = 0.35$ and background sound speed of $c_{s} = 3785\,\uvkms$, the corresponding characteristic flow velocity is $v'\approx 1350\,\uvkms$. This value results in a dynamical timescale on the energy injection scale of $t\approx 12\,\ut$, which is consistent with the assumed decay time of the forcing modes.
\section{Results and Discussion}\label{s:results}
In order to probe the behavior of model turbulence, we constructed a database of 2D and 3D models. Table \ref{t:models}
%
%
\ctable[
    cap = model binary systems table,
    caption = { Parameters and main properties of $512^{3}$ resolution models.},
    label = t:models,
    nostar
    ]
    {l c c c c}
    {
       \tnote[a]{Injection rate of turbulence forcing kinetic energy.}
       \tnote[b]{Root-mean-square turbulence Mach number averaged averaged over one turbulent turnover time before the end of the transient phase.}
       \tnote[c]{Simulated time of carbon ignition.}
       \tnote[d]{Simulated time of oxygen ignition.}
     }
    {\FL
    Model        &  $E_{k,td}$\tmark[a]   &  $\Mrms$\tmark[b]  &  $t_{C,ign}$\tmark[c]  &  $t_{O,ign}$\tmark[d]  \NN
    designation  &  ($\uEktd$)            &                    &  ($\ut$)               &  ($\ut$)               \ML
    H11          &  \num{1.1e15}          &  0.358             &  51.7                  &  56.1                  \NN 
    H13          &  \num{1.3e15}          &  0.377             &  45.8                  &  50.7                  \LL 
    }
presents the parameters and main properties of two high-resolution models obtained in this study (supporting several 2D models and 17 lower resolution 3D models are not listed). These models differ in terms of the amount of kinetic energy used to sustain turbulence and the resulting model turbulence Mach number, $\Mrms$. In our high-resolution models, H11 and H13, $\Mrms$ reached about 0.36 and 0.38, respectively, in agreement with the values obtained in lower resolution models. 

\begin{figure}
 \centering
 \includegraphics[width=\columnwidth]{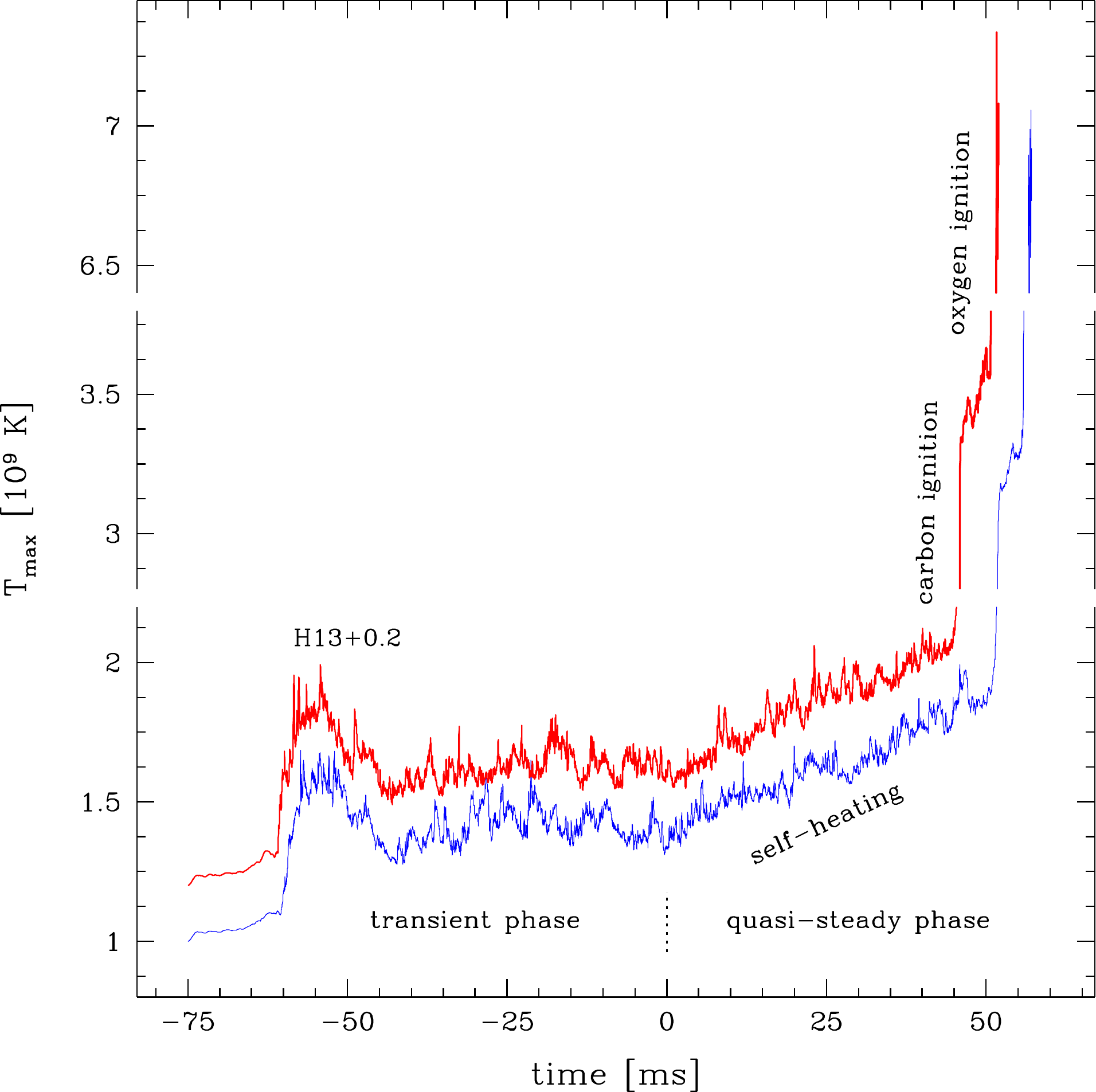}
 \caption{Evolution of maximum temperature in high-resolution models. The temperature is shown for model H11 and model H13 with a thin (blue in color version) and thick (red in color version) solid line, respectively. Note the temperature remains roughly constant by the end of the transient phase ($t=0\,$ms) and on average gradually increases after that time due to self-heating and the dissipation of turbulent kinetic energy. In both models, the temperature sharply rises in two stages, first when carbon is ignited and then again when oxygen detonates. Note the temperature scale is broken into three separate segments and H13 data is offset to improve readability.}
 \label{f:Tmax}
\end{figure}

After the quasi-steady state was achieved, nuclear burning was enabled at $t=0\ \ut$. Fig.\ \ref{f:Tmax}
shows the evolution of maximum temperature in the high resolution models. We found that the mean maximum model temperature gradually increased with time due to self-heating and dissipation of turbulent kinetic energy. At the same time, it fluctuated with a typical amplitude of up to about $5$ per cent around its mean value. The rapid increase in the maximum model temperatures seen around $t\approx45-50\ \ut$ marks the moment of carbon ignition, with the mean mass-weighted temperature for both models of $\approx\es{1.6}{9}\,\uT$.
\subsection{Birth and Growth of Carbon Deflagrations}\label{s:Cdef}
\begin{figure*}
    \centering
       \includegraphics[width=0.33\linewidth]{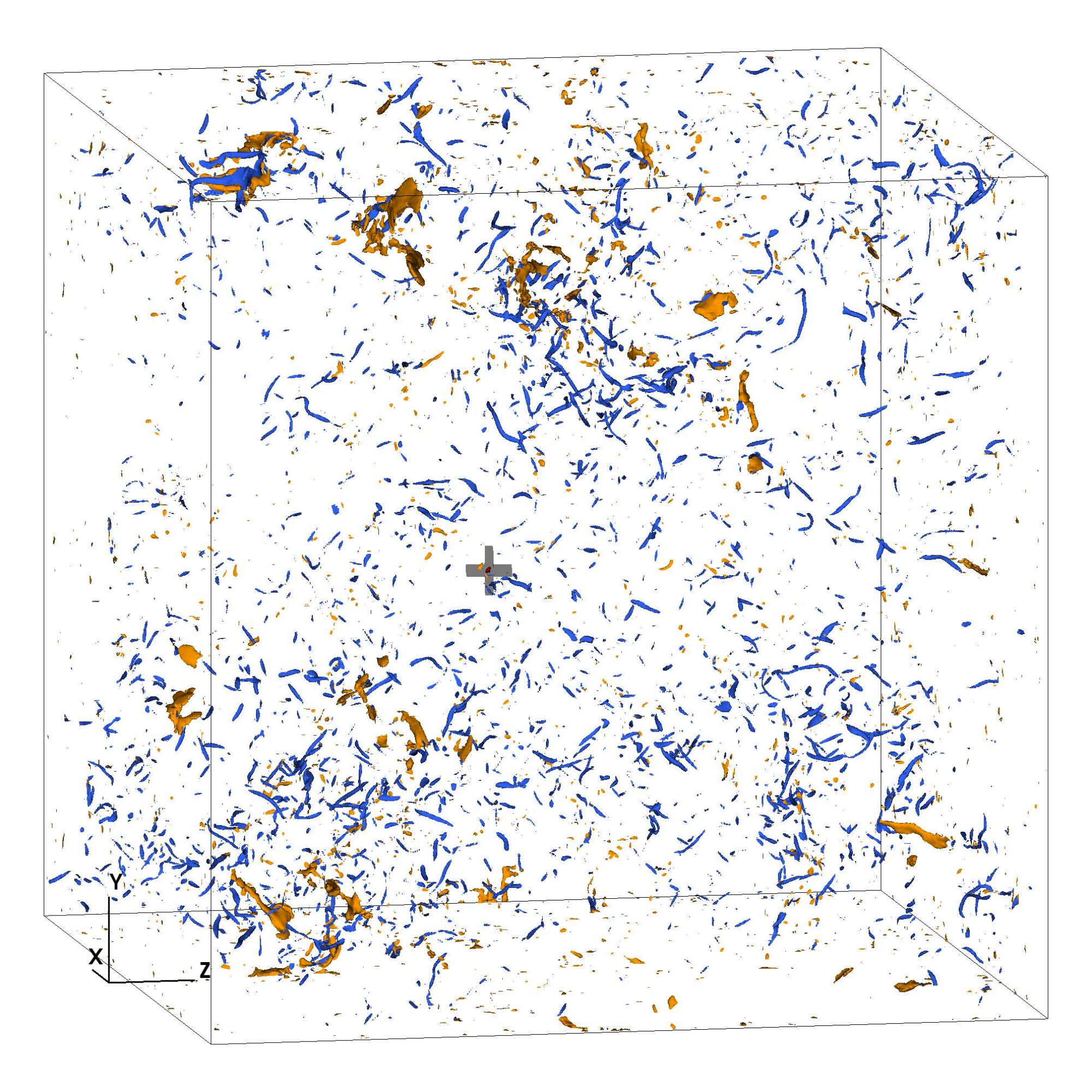} 
    \hfill
       \includegraphics[width=0.33\linewidth]{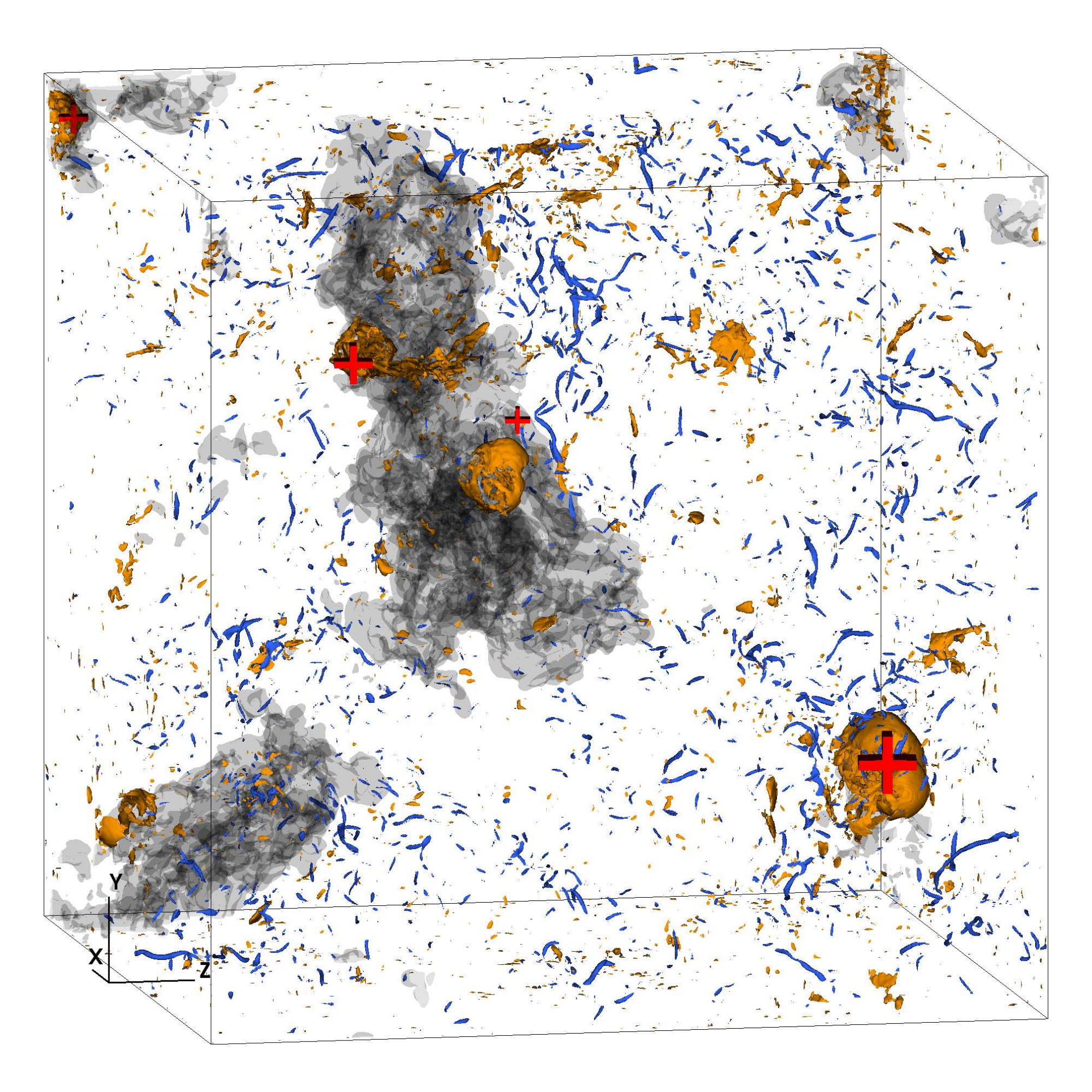} 
    \hfill
       \includegraphics[width=0.33\linewidth]{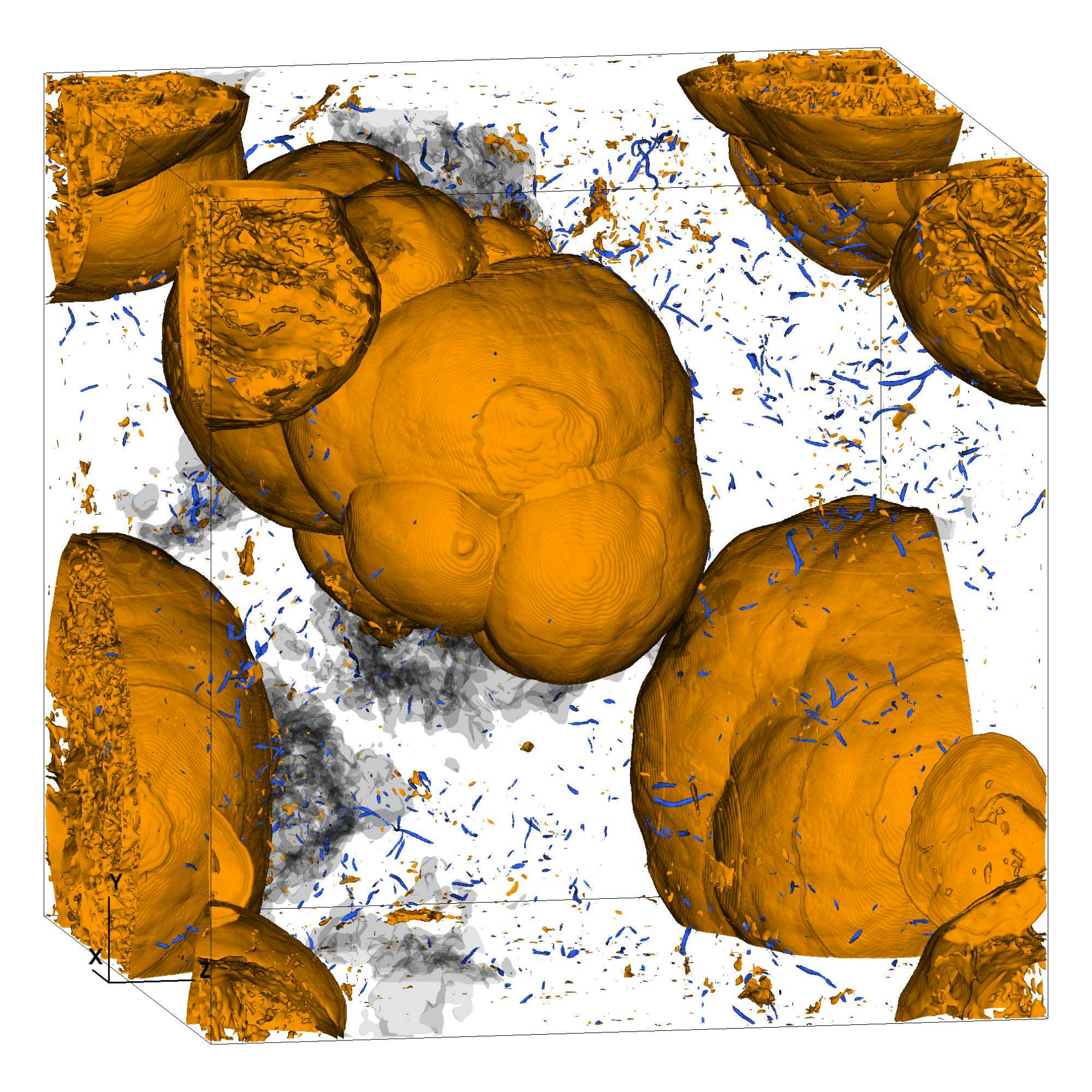} 
    \caption{DDT in the H11 model. Divergence and enstrophy are shown with black (yellow in color version) and gray (blue in color version) flooded contours, respectively. Strong compression is associated with shocklets (patch-like objects) while mixing is the most intense in regions occupied by vortex tubes (elongated objects). Contour levels correspond to the same values in all panels.
    (left panel; $t\approx 51.7$\,ms) The first carbon ignition region is marked with a cross symbol near the center of the computational domain. Note a relative absence of shocklets and vortex tubes in this region indicative of carbon ignition occurring in regions of smooth flow with weak mixing.
    (middle panel; $t\approx 56.1$\,ms) The first oxygen detonation region is marked with the largest cross symbol. Three select future oxygen detonation regions are also marked with smaller cross symbols; the smaller cross size indicates the longer time delay of oxygen detonation. Large amorphous structures are ashes produced by carbon deflagrations.
    (right panel; $t\approx 56.9$\,ms) The interacting oxygen detonations at the final simulation time.}
    \label{f:3Dmorphology}
\end{figure*}

In the left-hand panel of Fig.\ \ref{f:3Dmorphology}
we show the structure of the divergence and vorticity fields in model H11 at $t\approx 51.7$\,ms into the quasi-steady phase of the evolution and just prior to carbon ignition (an approximate location of the ignition point is marked with a cross symbol in the figure). At this time, the turbulent flow field is populated with randomly located vortex tubes with occasionally present segments of weak shock fronts. However, the ignition region is not directly associated with compression while the amount of vorticity in that region seems low. Instead, we found mild ignitions forming in regions with temperature elevated by superposition of acoustic perturbations and productive self-heating thanks to inefficient mixing (as evidenced by the relatively low values of vorticity) and lack of adiabatic cooling (thanks to strong plasma degeneracy; for typical density and temperature, $(\rho,T)_\mathrm{Cign} \approx (\es{1}{7}\ \urho,\ \es{1.6}{9}\ \uT)$, thermal contribution to the pressure is about $25$ per cent).

Over the next few milliseconds of simulated time, we observed a series of carbon ignitions taking place at several well-separated locations (middle panel in Fig.\ \ref{f:3Dmorphology}). Each such ignition event was accompanied by emission of a relatively weak acoustic signal. As time progressed, individual deflagrations grew in size making the flow increasingly inhomogeneous. Although we did not observe carbon deflagrations to directly produce detonations at this time in the evolution, carbon ignitions contributed to the acoustic background as evidenced by emission of weak acoustic waves from the ignition points. Therefore, this phase of the evolution might be considered as an early stage of the SWACER mechanism \citep{LeeYoshikawa1978}. Here temporal coherence of ignitions is on average improved by self-heating and modulated by acoustic perturbations contributed by previous ignition events.
\subsection{Initiation and Role of Transient Carbon Detonations}\label{s:failedCdet}
We observed the formation of carbon detonations a few milliseconds into the carbon deflagration phase. These detonations can be seen as strongly compressed, nearly spherical regions located near the three smallest crosses in the middle panel in Fig.\ \ref{f:3Dmorphology}. They were initiated by intermittently present, strong \emph{shocklets created by turbulence}. Upon entering highly convoluted regions of partially burned material, these shocklets were able to acoustically couple with burning (inside carbon-rich pockets) and detonate carbon as a result. Therefore, these detonations were a direct consequence of turbulence becoming strongly inhomogeneous at this intermediate time.

Because these carbon detonations were formed inside regions occupied by deflagrating material, they quickly weakened after entering carbon-depleted fuel. Furthermore, these detonations were not supported in their evolution toward oxygen detonations as reactivity of the unburned material remained at low at this time. As we will see later (see Section \ref{s:Cdet}), suitable preconditioning of the environment is indeed required for oxygen to ignite. Despite their disappointing fate, these early detonations nevertheless played an important role in the evolution toward the supernova explosion. This is because the acoustic signal that they produced was much stronger than that of the perturbations created by mild carbon ignitions. Therefore, these transient carbon detonations constituted one of the key components of the participating SWACER mechanism.
\subsection{Zel'dovich-driven Carbon Detonations}\label{s:Cdet}
At a later time, the maximum temperature continued to fluctuate, but now with reference to the temperature of the carbon ash of $\approx \es{3.3}{9}\ \uT$. The second rapid increase in the maximum temperature seen in Fig.\ \ref{f:Tmax} around $50\ \ut$ into the quasi-steady phase marks the time of oxygen ignition. We studied this process in detail in a series of 2-dimensional models with resolution systematically increased to slightly less than one meter. (The corresponding mesh dynamic range in the best resolved model was $32,768$.) We found that only at that level of resolution can we truthfully describe the gradual transition from mild carbon ignition to carbon detonation.

Although the initial stages of explosion involved transient carbon detonations (cf.\ Section \ref{s:failedCdet}), the carbon detonation described here had a sustained character. This difference was due to naturally occurring \emph{preconditioning} of the cold fuel. Note, that during about the $5\ \ut$ that elapsed after the first deflagration was born, several more deflagrations were produced. Even though in this work we do not model the flame, deflagrations gradually grew in size in our simulations due to numerical mixing between the hot carbon ash and the cold fuel. We cannot claim the observed flame growth to be realistic, but it is conceivable that a qualitatively similar effect occurs in nature.

\begin{figure*}
    \centering
        \includegraphics[width=0.33\linewidth]{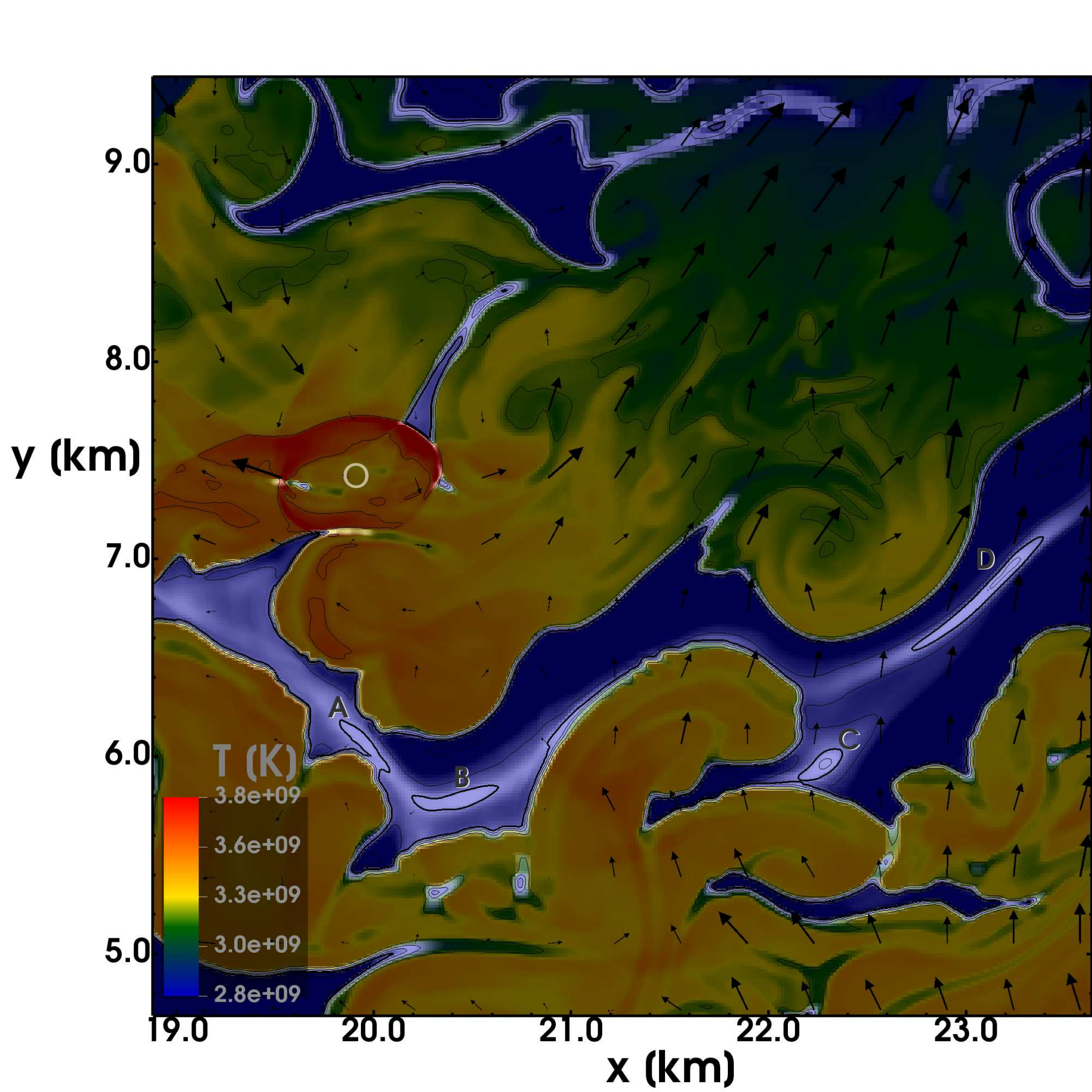}
    \hfill
        \includegraphics[width=0.33\linewidth]{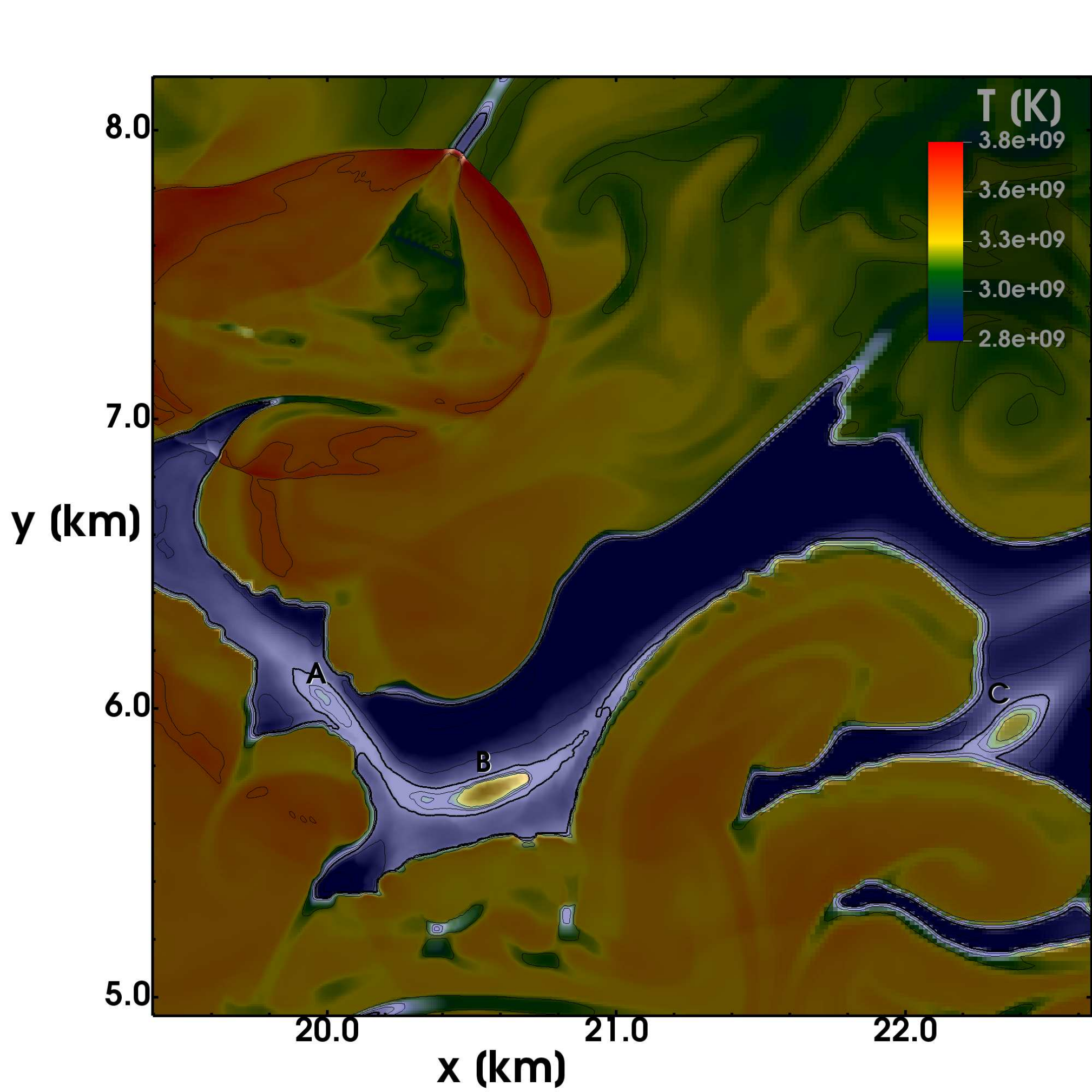}
    \hfill
        \includegraphics[width=0.33\linewidth]{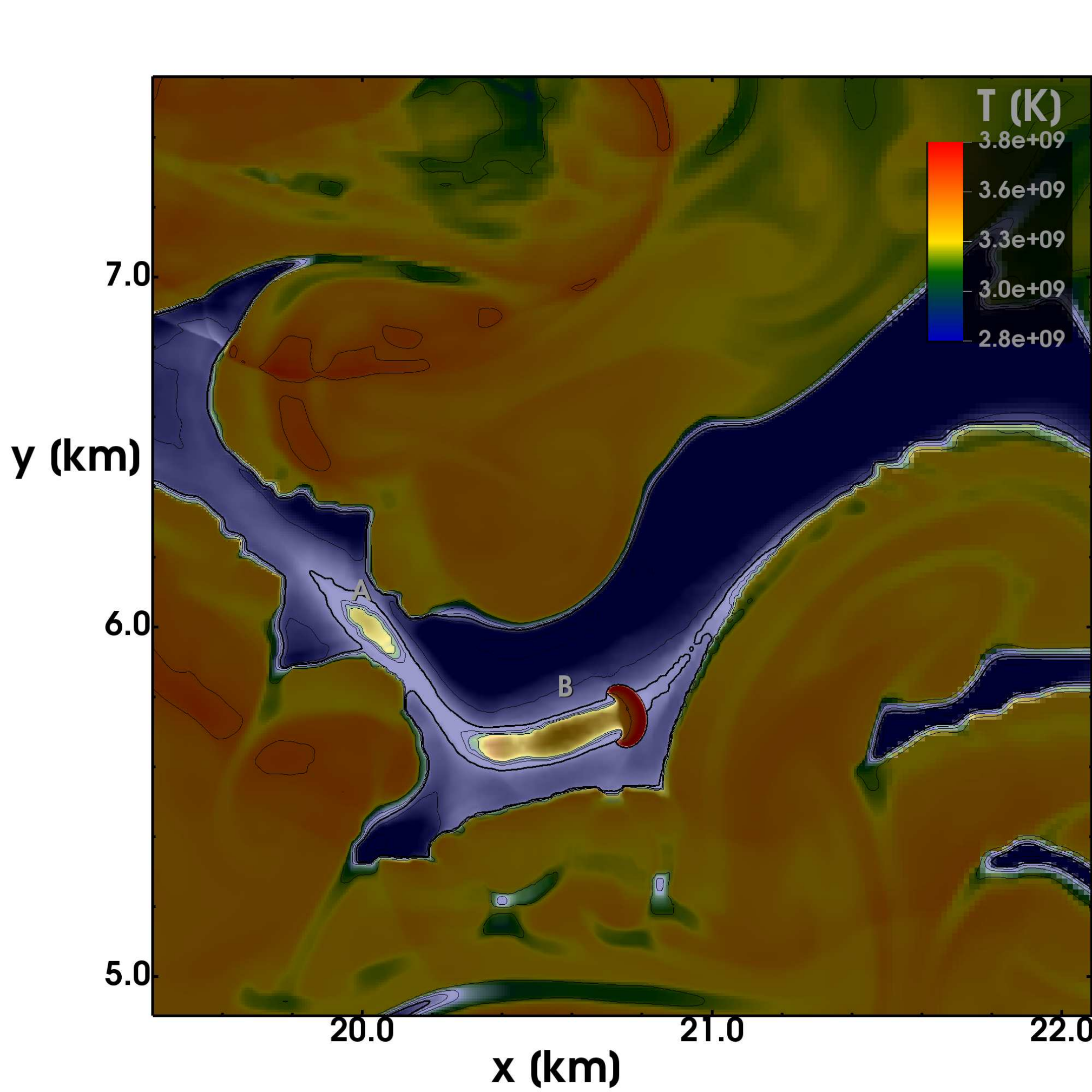}
    \caption{Process of initiation of a sustained carbon detonation in a 2-dimensional model of turbulent, dense stellar plasma.
    (left panel; reference time) Several prospective detonation kernels are marked along a channel filled with a fuel of increased reactivity (shortened ignition time).
    (middle panel; 335 $\mu$s time offset) Intense self-heating leads to the formation of an acoustic perturbation due to thermal expansion of kernel B.
    (right panel; 535 $\mu$s time offset) A nascent carbon detonation runs into the fuel with its central segment steered by the reactivity gradient.
    In the panels, temperature is shown with pseudocolor maps. The color scale is saturated below $2.8\times 10^9$\,K, with the lowest temperatures reaching about $1.4\times 10^9$\,K. There are 7 equally spaced temperature contour lines between $1.45-3.55\times 10^9$\,K. A thick temperature contour line corresponds to a temperature of $2.15\times 10^9$\,K, which roughly corresponds to the carbon ignition temperature. In the left-hand panel, an open circle marks a location of transient carbon detonation; the velocity field in a reference frame comoving with kernel B is shown with arrows with the maximum arrow length corresponding to $\approx 3300$ km\,s$^{-1}$. Note the spatial scale changes between the panels. See text and Fig.\ \ref{f:lineouts} for details.
    }
    \label{f:Cdet}
\end{figure*}

The presence of turbulent perturbations and active burning, as described in Section \ref{s:Cdef}, had profound consequences for the evolution of the remaining cold fuel. In the left-hand panel of Fig.\ \ref{f:Cdet},
we show the morphology and dynamics of a partially deflagrating region of the 2-dimensional model. As one can note, the narrow (width $\approx 500\ \mathrm{m}$), meandering channel of fuel trapped between chunks of carbon deflagrations stretches from $(x,y) \approx (19,6.8)\ \mathrm{km}$ to $(x,y) \approx (23.5,7.2)\ \mathrm{km}$. The channel material displays variation in the temperature and the related variations in the ignition time (reactivity), with color brightness indicating regions with higher reactivity. The variations take the form of elongated structures generally aligned with the channel shape. They were created by \emph{adiabatic compression} on the scale of resolved turbulent eddies, which is several times larger than the channel width. 

The presence of such perturbations is evident after examining the velocity field. We note the existence of a velocity gradient along the direction from $(x,y) \approx (19,9.2)\ \mathrm{km}$ to $(x,y) \approx (22,5.3)\ \mathrm{km}$ in the left-hand panel of Fig.\ \ref{f:Cdet}. This velocity gradient compressed the channel material and produced a prominent maximum of reactivity near $(x,y) \approx (20.5,5.8)\ \mathrm{km}$ (see region labelled B in the figure). We also note the presence of additional maxima of reactivity, regions labelled A, C, and D in the figure. It is conceivable that additional compression might be due the potential velocity perturbations induced by thermal expanding deflagrations \citep{Sabelnikov2019}. This effect could be important especially in narrow segments of fuel channels (e.g.\ near region A). In any event, rapid and spatially localized increase in plasma reactivity is a direct consequence of strong plasma degeneracy as the fuel temperature is allowed to increase under nearly constant density.

\begin{figure}
 \includegraphics[width=0.995\columnwidth]{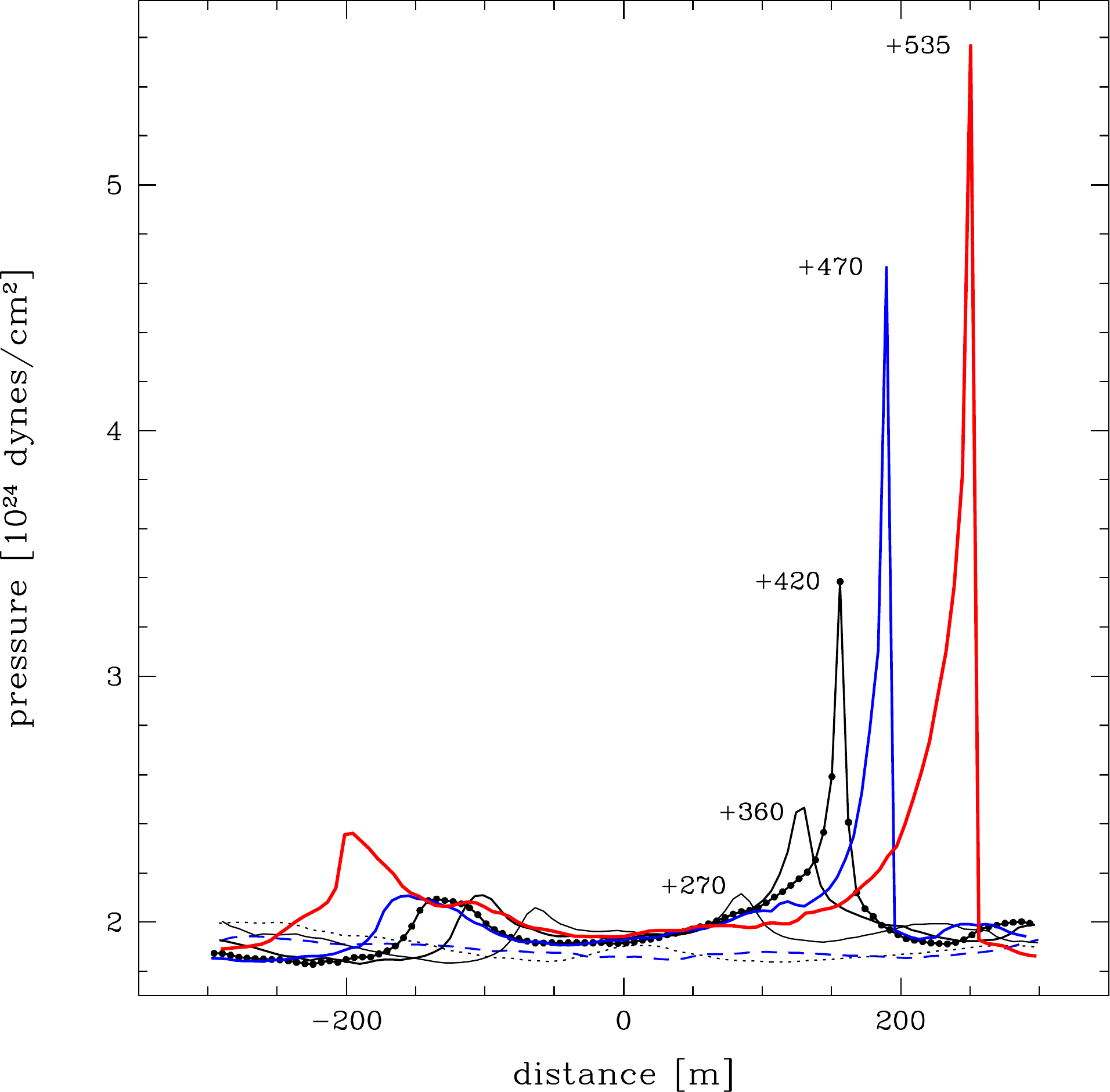}
 \caption{Evolution of pressure along the mid-section of kernel B (see Figure \ref{f:Cdet}). Pressure profiles are shown at the reference time just prior to carbon ignition (dashed line), after 136 $\mu$s (dotted line), and at later times are labeled with corresponding time offsets. Individual mesh cell data are shown for time offset $420$ $\mu$s with full circles to illustrate mesh resolution ($\approx 100$\,cm in the model) and the relative steepness of a rapidly strengthening detonation.}
 \label{f:lineouts}
\end{figure}

The evolution of region B is of primary interest in the subsequent evolution. Fig.\ \ref{f:lineouts}
shows a series of pressure line-outs centered at the midsection of this region, and extending over $\approx 500\ \mathrm{m}$ \citep[cf.\ Fig. 3 in] {Khokhlov+97}. Initially (reference time; shown with a dashed line in Fig.\ \ref{f:lineouts}), the pressure in this region is smooth and nearly constant. After $13\,6\mu$s (dotted line in Fig.\ \ref{f:lineouts}), the pressure shows elevated values over the central $\approx 100\ \mathrm{m}$-long section. Thanks to continuing energy deposition, the local nuclear burning time eventually becomes shorter than the mesh cell sound-crossing time marking the onset of detonation initiation \citep{Glasner+2018}. At that time, a weak acoustic perturbation is formed (middle panel in Fig.\ \ref{f:Cdet}). As the self-heating accelerates, the plasma degeneracy is lifted and the energized perturbation rapidly steepens into a shock. Because the reaction zone remains acoustically coupled to the wave at all times, a fully fledged carbon detonation is eventually born (pressure profile at time offset of 535\,$\mu$s in Fig.\ \ref{f:lineouts}). The leading segment of this detonation front continues to expand steered in its evolution by the reactivity gradient, however, the wave is sufficiently strong to ignite carbon also in the surrounding fuel (cf.\ the right-hand panel in Fig.\ \ref{f:Cdet}). 

The described mechanism bears all the signatures of the Zel'dovich-Neumann-D\"oring detonation model \citep{FickettDavis1979}. The reason why earlier carbon detonations were transient (cf.\ Section \ref{s:failedCdet}) was the lack of a suitably preconditioned environment. Such an environment is a product of inhomogeneous turbulence. It consists of extended regions of cold fuel which are subject to adiabatic heating on a scale of resolved turbulent eddies. This process is further aided by acoustic energy contributed by preceding events of mild carbon ignition and of transient carbon detonations, as postulated in the SWACER model \citep{LeeYoshikawa1978}. Because early shocklet-initiated carbon detonations were found to fail in the absence of suitably preconditioned regions, we conclude that preconditioning is, indeed, a \emph{necessary} condition for successful carbon detonation to occur.
\subsection{Paths Toward Oxygen Detonations}
We observed oxygen detonations to form when an inert shock produced by a carbon detonation exits carbon ash and enters a preconditioned region. Such preconditioned regions are associated with cusps of "carbon flame" where fuel is compressed by thermal expansion of the partially burned material \citep[see][and Section \ref{s:Cdet}]{Sabelnikov2019}. This interaction results in carbon ignition and allows for the oxygen reaction zone to couple with the shock.

Also, we observed oxygen detonations due to the collision of carbon detonations. Such collisions were enabled by the highly unsteady evolution of the original carbon detonation through the cold fuel channel (Section \ref{s:Cdet}) and its subsequent propagation as a nearly inert shock through incompletely burned carbon ash. As various segments of the shock exited the deflagrating mass at nearby locations, carbon detonations were produced and their collision was sufficient to detonate oxygen. 
\section{Conclusions}\label{s:conclusions}
We studied the evolution of weakly compressible turbulence in carbon/oxygen mixtures to identify the necessary condition for a delayed detonation to occur in a single degenerate SN Ia formation channel. We found the DDT to occur naturally with a mild ignition of carbon followed by a two-stage detonation. Transient carbon detonations are initiated directly by shocks while sustained carbon detonations are initiated with help of the Zel'dovich reactivity gradient mechanism induced by compressive turbulence modes on intermediate scales. The subsequent oxygen detonations are due to either gradual strengthening of a nearly inert shock by reactivity gradient, or interaction between carbon detonations as they emerge from the partially burned material. The evolution toward DDT is accelerated by the SWACER mechanism.

Turbulent mixing and interaction between different types of detonations with the inhomogeneous fuel may produce observational signatures helpful in validating the proposed explosion model. In particular, the results hint at a complete carbon burn but leave a possibility for oxygen enhancement or an incomplete oxygen burn.

We found the model DDT timescale increasing as the turbulent intensity decreases. Because in a centrally ignited scenario stellar layers with density $\approx 10^7$\,g cm$^{-3}$ expand on a timescale $\approx 150$\,ms before the density becomes too low for ignition, we do not expect DDT to occur when turbulence is too weak. On the other hand, some intensification of pre-existing turbulence might be possible due to compression caused by the rising flame plumes. These possibilities lend support to future studies of structure and evolution of outer layers of accreting massive white dwarfs.

\section*{Acknowledgements}
This research used resources of the National Energy Research Scientific Computing Center, which is supported by the Office of Science of the U.S.\ Department of Energy under Contract No.\ DE-AC02-05CH11231. The software used in this work was in part developed by the DOE Flash Center at the University of Chicago. Data visualization was performed in part using VisIt \citep{HPV:VisIt}. This research has made use of NASA's Astrophysics Data System Bibliographic Services.
\section*{Data availability}
The data underlying this article will be shared on reasonable request to the corresponding author.
%
%
%
%
%
%
\bibliographystyle{mnras}
\bibliography{citations}
%
%
%
\label{lastpage}
\end{document}

%% file: packageIncludes.tex
\usepackage{amssymb}
\usepackage{fp}
\usepackage{xfrac}
\usepackage{amsmath}
\usepackage{booktabs}
\usepackage{cancel}
\usepackage{colortbl}
\usepackage{csquotes}
\usepackage{datatool}
\usepackage{helvet}
\usepackage{mathpazo}
\usepackage{listings}
\usepackage{multirow}
\usepackage{array}
\usepackage{float}

\usepackage{siunitx}

\usepackage{xparse}
\usepackage{etoolbox}
\usepackage{isotope}
\usepackage{ctable}
\usepackage{url}

\usepackage{graphicx}

%% file: figureMacros.tex
\newlength\figureheight 
\newlength\figurewidth 
